\documentclass[entropy,article,submit,moreauthors,10pt,a4paper]{mdpi} 

\usepackage{graphicx}
\usepackage{subfigure}
\usepackage{amsmath}
\usepackage{amssymb}
\newcommand{\ket}[1]{\left\vert#1\right\rangle}
\newcommand{\bra}[1]{\left\langle#1\right\vert}

\newcommand{\tr}{\mathrm{tr}}
\firstpage{1} 
\articlenumber{x}
\doinum{10.3390/------}
\pubvolume{xx}
\pubyear{2016}
\copyrightyear{2016}
\externaleditor{Academic Editor: name}
\history{Received: date; Accepted: date; Published: date}

\Title{R{\'e}nyi divergences, Bures geometry and quantum statistical thermodynamics}

\Author{Ali \"{U}. C. Hardal, \"{O}zg\"{u}r E. M\"{u}stecapl{\i}o\u{g}lu}
\AuthorNames{Ali \"{U}. C. Hardal, \"{O}zg\"{u}r E. M\"{u}stecapl{\i}o\u{g}lu}

\address[1]{Department of Physics, Ko\c{c} University, \.Istanbul, 34450, Turkey}

\corres{Correspondence: ahardal@ku.edu.tr}

\abstract{The Bures geometry of quantum statistical thermodynamics at thermal equilibrium is investigated by introducing the connections between the Bures angle and the R{\'e}nyi $1/2$-divergence. Fundamental relations concerning free energy, moments of work and distance are established.}

\begin{document}

\section{Introduction}\label{sec:intro}
Galileo once said ``philosophy is written in the language of mathematics and the characters are triangles, circles and other figures"~\cite{galilei1979saggiatore}.  Since then, natural philosophy and geometry evolved side by side, leading groundbreaking perceptions of the foundations of todays modern science. A particular example introduced by Gibbs~\cite{gibbs1, gibbs2} who successfully described the theory of thermodynamics by using convex geometry in which the coordinates of this special phase space are nothing but the elements of classical thermodynamics.~While the Gibbsian interpretation has been successful for the understanding of the relations among the thermodynamical entities, it lacks the basic notion of a geometry: the distance. The latter constitutes the fundamental motivation for Weinhold's geometry~\cite{weinhold1975metric} where an axiomatic algorithm leads a scalar product and thus, induces a Riemannian metric into the convex Gibbsian state space. Later, Ruppeiner took the flag with the notion that any instrument of a geometry that describes a thermodynamical state of a system should lead physically meaningful results~\cite{ruppeiner1979thermodynamics}. To that end, the Riemannian scalar curvature of an embedded metric has been related with the thermodynamical criticality~\cite{ruppeiner1981application} and the the theory has been applied many classical as well as quantum mechanical systems~\cite{ruppeiner1995riemannian}. 

It is quite logical to search for abstract relations among geometry and thermodynamical processes, since that is what we are familiar with in the ``macro-world". In particular, the proportionality of the work done by a physical system during a transformation to the distance between start and end points provides strong motivation for the subject matter. Here, we explore if a similar relation exists within the theory of equilibrium quantum statistics. We do not over-complexify our strategy and indigenise the Descartian rule of thumb~\cite{descartes1637}: any problem in geometry can be reduced that of a problem of finding the lengths (the distances) of straight lines (between any given points). We show that the latter idea indeed leads to new insights concerning quantum statistical thermodynamics, if we measure the distance between two arbitrary quantum thermal states relative to an ``unbiased" reference state.

Such a reference state must be unique and must have geometrical as well as physical well defined interpretations. To that end, we make all our calculations with respect to the maximally mixed state which is the maximum entropy state from thermodynamical point of view and it lies in the geometrical centre of state space of density operators~\cite{bengtsson2006geometry}. We use the elements of the well known Bures geometry of the space of density operators~\cite{bengtsson2006geometry,bures1969extension}, i.e., we use the elements of the geometry of quantum statistical ensembles and search for physically meaningful relations concerning thermodynamics. The Bures distance has already been considered in quantum statistical mechanics~\cite{twamley1996bures}, non equilibrium thermodynamics~\cite{deffner2011nonequilibrium,deffner2013thermodynamic} and quantum phase transitions~\cite{zanardi2007bures,zanardi2007information,you2007fidelity}. Very recently, it has been shown to be a useful tool for determining the inner friction during thermodynamical processes, as well~\cite{plastina2014irreversible}.

While one can assume a more abstract mathematical path that may end up with similar expressions as found in this contribution, we bring another player to the gameplay that makes the thermodynamical connections more clear. We first show the relation between the Bures distance with the R{\'e}nyi $\alpha$-divergences~\cite{rrnyi1961measures,muller2013quantum} for a specific value of $\alpha=1/2$. The R{\'e}nyi divergences is shown to be the generalisations of the quantum entropies from which the measures of quantum information can be recovered~\cite{muller2013quantum}. Statistical thermodynamics based on Renyi divergenses has also been discussed in the literature~\cite{plastino1997universality, lenzi2000statistical, misra2015quantum} under strict constraints, i.e., maximazing the divergence itself with the assumption of a fixed internal energy and considering quasistatic isothermal processes with general interest in systems that are far from equilibrium. Second, the divergences has already played critical roles in the search for the laws of quantum thermodynamics~\cite{horodecki2013fundamental,brandao2015second,lostaglio2015description}. Finally, as we shall see here, the choice of $\alpha=1/2$ directly relate the curved geometry of space of the density operators to the quantum entropies as well as it has the physical meaning of being the measure of maximal conditional entropy between the statistical ensembles under consideration~\cite{muller2013quantum}. 

Equipped with the R{\'e}nyi divergence, we, first, establish fundamental identities among distance and occupation probabilities of a given quantum thermal state. Later on, we use same identities to further develop our approach to unearth the implicit connections between distance, free energy and work distribution during thermal transformations that occurs between equilibrium quantum states.

This paper is organised as follows. In Section~\ref{sec:density}, we introduce the necessary elements of Bures geometry, R{\'e}nyi divergences and their relations. In Section~\ref{sec:results}, we present our results. We conclude in Section~\ref{sec:conc}.
\section{The space of density operators}\label{sec:density}
A complex $N\times N$ matrix $\rho$ acting on a Hilbert space $\mathbb{H}$ of dimension $N$ is called a density matrix if it is positive semi-definite ($\bra{\psi}\rho\ket{\psi}\geq0$ $\forall \ket{\psi}\in\mathbb{H}$), Hermitian ($\rho=\rho^{\dagger}$) and normalized ($\tr{\rho}=1$). The set of density matrices, $\mathbb{D}$, is the intersection of the space of all positive operators $\mathbb{P}$ with a hyperplane parallel to the linear subspace of traceless operators~\cite{bengtsson2006geometry}. $\mathbb{D}$ is a convex set and the maximally mixed state
\begin{equation}\label{eq:mmx}
\rho_{*}=\frac{1}{N}\mathbf{1}_{N},
\end{equation}
lie on its centre with $\mathbf{1}_{N}$ being the $N\times N$ identity operator.
\subsection{The Bures geometry}
There exists a family of monotone metrics~\cite{bengtsson2006geometry} that can be used to measure the geometrical (statistical) distance between any given two density operators $\rho$, $\sigma$ $\in\mathbb{D}$. Among all of these measures, the minimal one is given by the Bures distance~\cite{bures1969extension}
\begin{equation}\label{eq:bures}
D_{B}(\rho,\sigma) = \tr{\rho}+\tr{\sigma}-2F(\rho,\sigma),
\end{equation} 
where $F(\rho,\sigma)=\tr\sqrt{\sqrt{\sigma}\rho\sqrt{\sigma}}$ is the Uhlmann's root fidelity~\cite{uhlmann1992metric}. The Bures distance measures the length of the curve between $\rho$ and $\sigma$ within the set of all positive operators $\mathbb{P}$, while the length of the curve within $\mathbb{D}$ is measured by the Bures angle~\cite{uhlmann1992metric,uhlmann1993density,uhlmann1996spheres}
\begin{equation}\label{eq:angle}
\cos{d_{B}}(\rho,\sigma)=F(\rho,\sigma), \hspace{4pt}0\leq d_B\leq\frac{\pi}{2}.
\end{equation}
There is a Riemannian metric, the Bures metric, associated with the distances (\ref{eq:bures}) and (\ref{eq:angle}). First, we recall that, due to the positivity property, any density operator $\rho\in\mathbb{D}$ acting on $\mathbb{H}$ can be written as $\rho=KK^{\dagger}$ where $K:=\sqrt{\rho}U$ with $U$ being a unitary operator. We can search for a curve $\rho(\tau)$ within       
$\mathbb{D}$ by imposing that the amplitudes remain parallel 
\begin{equation}\label{eq:parallel}
\dot{K}^{\dagger}K=K^{\dagger}\dot{K},
\end{equation}
where $\dot{K}$ denotes the differentiation of $K$ with respect to the arbitrary parametrization $\tau$. The condition (\ref{eq:parallel}) is satisfied if $\dot{K}=GK$~\cite{uhlmann1989berry} where $G$ is an Hermitian matrix. It follows that
\begin{equation}\label{eq:rho_evo}
d\rho=G\rho+\rho G.
\end{equation}
The Bures metric $ds_{B}^2$ is then defined as
\begin{equation}\label{eq:bures_metric}
ds_{B}^{2}=\frac{1}{2}\tr{Gd\rho}.
\end{equation}
If the density operator is strictly positive $\rho>0$, the matrix $G$ is uniquely determined. Note that, although there are some methods to determine the Bures metric exactly~\cite{twamley1996bures}, it is generally very hard to compute and exact form of this metric will not be needed in the current contribution.  

The elements of the Bures geometry has been used in the context of quantum statistics in the literature. In particular, the Bures metric (and the distances that belong to it) proven to be useful tools for the characterisation of quantum criticality and quantum phase transitions~\cite{zanardi2007bures,zanardi2007information,you2007fidelity}. The Bures distance is proven to be a lower bound for the estimation of the energy-time uncertainty, as well~\cite{uhlmann1992energy, uhlmann2009geometry}. Another neat example provided by Twamley~\cite{twamley1996bures} where the Bures distance between two squeezed thermal states evaluated and the curvature of the corresponding Bures metric suggested as a measure to optimize detection statistics. More resent studies revealed the deeper connections between the quantum thermodynamical processes and the Bures geometry. Specifically, it has been shown that the quantum irreversible work~\cite{deffner2011nonequilibrium,deffner2013thermodynamic} as well as the quantum inner friction~\cite{plastina2014irreversible} bounded from below by the Bures angle.

Here, we look at the picture from a different perspective. We shall calculate the distance between two equilibrium quantum thermal states with respect to the maximally mixed state. It turns out that this relative distance can be written in terms of the difference between the corresponding free energies, leading to fundamental relations between work, distance and efficiency. But first, we should equip ourselves with the R{\'e}nyi divergences to make the statistical connections more clear.
\subsection{The R{\'e}nyi $\alpha$-divergences}
The quantum R{\'e}nyi $\alpha$-divergences~\cite{muller2013quantum} are the generalizations of the family of Renyi entropies~\cite{rrnyi1961measures} from which the measures of quantum information can be recovered. Moreover, the R{\'e}nyi divergences admit a central role in so-called generalized quantum second laws of thermodynamics~\cite{horodecki2013fundamental,brandao2015second} as well as laws of quantum coherence that enters thermodynamical processes~\cite{lostaglio2015description}. In their most general form, for two given density operators $\rho$, $\sigma$ $\in\mathbb{D}$, the $\alpha$-divergences defined as~\cite{muller2013quantum,mosonyi2014quantum}
\begin{equation}\label{eq:renyi_div}
  S_{\alpha}(\rho||\sigma)=\begin{cases}
    \frac{1}{\alpha-1}\ln\tr\rho^{\alpha}\sigma^{1-\alpha}, & \alpha\in[0,1).\\
    \frac{1}{\alpha-1}\ln\tr(\sigma^{1-\alpha/2\alpha}\rho\sigma^{1-\alpha/2\alpha})^{\alpha}, & \alpha>1.
  \end{cases}
\end{equation}
Here, we remain in the domain of equilibrium statistical thermodynamics at finite temperature and to do so we shall only deal with a single preferred value of $\alpha=1/2$. Physically, it corresponds the maximum conditional entropy between the states $\rho$ and $\sigma$~\cite{muller2013quantum} and the divergence reads
\begin{equation}\label{eq:renyi_one_two}
S_{1/2}(\rho||\sigma)=-2\ln\tr\sqrt{\rho}\sqrt{\sigma}.
\end{equation}
It is easy to see that the argument of the logarithm in the right hand side of Eq.~(\ref{eq:renyi_one_two}) is nothing but the root fidelity $F(\rho,\sigma)$ for two commuting operators $\rho$, $\sigma$ $\in\mathbb{D}$. Thus, a natural relation between the entropy function and the Bures geometry is constructed to give 
\begin{equation}\label{eq:renyi_bures}
S_{1/2}(\rho||\sigma)=-\ln\cos^2{d_B(\rho,\sigma)}, \hspace{4pt}[\rho,\sigma]=0, \hspace{4pt}0\leq d_B\leq\frac{\pi}{2}.
\end{equation}
Note that, here we follow Ref.~\cite{mosonyi2014quantum} for the definition of R{\'e}nyi divergences~(\ref{eq:renyi_div}). In Ref.~\cite{muller2013quantum}, divergences presented as $d_{\alpha}\equiv[1/(\alpha-1)]\ln\tr\rho^{\alpha}\sigma^{1-\alpha}$ for $\alpha\in (0,1)\cap (1,2]$ and $d_{\alpha}^{\prime}\equiv[1/(\alpha-1)]\ln\tr(\sigma^{(1-\alpha)/(2\alpha)}\rho\sigma^{(1-\alpha)/(2\alpha)})^{\alpha}$ for $\alpha\in[1/2,1)\cap(1,\infty)$. The authors did notice the relation $d_{1/2}^{\prime}=-2\ln F(\rho,\sigma)$ but also write $d_{1/2}^{\prime}\neq d_{1/2}$. Here, we state indeed $d_{1/2}^{\prime}= d_{1/2}$ if $[\rho,\sigma]=0$. More, the commutativity condition provides consistency to our theory due to the fact that we shall only be dealt with thermal states in the energy eigenbasis where they are diagonal. In this case, in fact, the Bures angle~(\ref{eq:angle}) and the Bures distance~(\ref{eq:bures}) are equivalent to the classical Bhattacharyya and Hellinger statistical distances~\cite{bengtsson2006geometry}, respectively. The latter connections signifies the statistical significance of the divergence $S_{1/2}(\rho||\sigma)$.  
 
The relation~(\ref{eq:renyi_one_two}) will be the starting point of our interpretation of equilibrium geometric thermodynamics in the next section with the final ingredient of a suitable reference point~(\ref{eq:mmx}). 
\section{Results}\label{sec:results}
Here, we present our contributions to the geometric interpretations of quantum statistical thermodynamics. We shall start with fundamental relations among equilibrium fluctuations, Renyi divergences and Bures angle. Afterwards, we shall provide a relation concerning the distance between two equilibrium quantum states and the change in the corresponding free energies. We finalize this section by prividing a relation between work, distance and Carnot efficiency.
\subsection{Fundamental relations}
Let $\rho_{th}$ be a thermal state of quantum mechanical system, acting on a $N$-dimensional Hilbert space and described by the Hamiltonian $\mathrm{H}$ at finite equilibrium temperature $\beta=1/k_B T$ with $k_B$ being the Boltzmann constant. We have
\begin{equation}\label{eq:thermal_general}
\rho_{th}=\sum_{i=1}^{N}p_{i}\ket{\psi_i}\bra{\psi_i},
\end{equation}
where $p_i = e^{-\beta E_i}/Z$ are the occupation probabilities, $\ket{\psi_i}$ and $E_i$ are the eigenvectors and the corresponding eigenvalues of the Hamiltonian $\mathrm{H}$ that satisfy the eigenvalue equation $\mathrm{H}\ket{\psi_i}=E_i\ket{\psi_i}$ and $Z=\tr{e^{-\beta\mathrm{H}}}$ is the partition function. 

The R{\'e}nyi divergence of $\rho_{th}$ with respect to the maximally mixed state $\rho_{*}$ becomes
\begin{equation}\label{eq:fund1}
S(\rho_{th}||\rho_{*})=-2\ln F(\rho_{th},\rho_{*})=-2\ln\frac{1}{\sqrt{N}}\sum_{i=1}^{N}p_i^{1/2},
\end{equation}
where we set $S_{1/2}(\rho||\sigma)\equiv S(\rho||\sigma)$ for the sake of simplicity. We obtain
\begin{equation}\label{eq:fund2}
\sum_{i=1}^{N}p_i^{1/2}=\sqrt{N}e^{-S(\rho_{th}||\rho_{*})/2}.
\end{equation}
First immediate consequence of Eq.~(\ref{eq:fund2}) is that since $p_i\leq p_{i}^{1/2}$ $\forall i$, we have $\sum_i p_{i}^{1/2}\geq 1$ and thus $\ln N\geq S(\rho_{th}||\rho_{*})$ as expected. 

We continue by taking the square of both sides of the fundamental equation~(\ref{eq:fund2}) to obtain
\begin{equation}\label{eq:fund5}
1+2\sum_{i<j}^{N}p_{i}^{1/2}p_{j}^{1/2}=Ne^{-S(\rho_{th}||\rho_{*})}.
\end{equation}
Inspired by the Wootter's statistical distance~\cite{wootters1981statistical}, we may introduce a distinguishability measure between the energy eigenstates of the system that is given by
\begin{equation}\label{eq:fund6}
\cos{d_W}=\sum_{i<j}^{N}p_{i}^{1/2}p_{j}^{1/2},
\end{equation}
which is a function of fidelity through Eq.~(\ref{eq:fund1}).

Finally, using Eq.~(\ref{eq:renyi_bures}) and Eq.~(\ref{eq:fund6}), we arrive at the relation
\begin{equation}\label{eq:fund7}
\cos^2{d_{B}}=\frac{1+2\cos{d_W}}{N}.
\end{equation}
The strict positivity of $\cos{d_W}$ requires $0\leq d_B<\pi/2$ which leads $(N-1)/2\geq\cos{d_W}>0$. 
\subsection{Geometry, entropy and the thermodynamical free energy}
We start with rewriting Eq.~(\ref{eq:fund1}) as
\begin{equation}\label{eq:fund1_1}
S(\rho_{th}||\rho_{*})=\ln{N}+\ln{Z}-2\ln{Z^{\prime}},
\end{equation}
where $Z=\sum_{n}e^{-\beta E_{n}}$ and $Z^{\prime}=\sum_{n}e^{-\beta E_{n}/2}$. Let $\rho^{(1)}$ and $\rho^{(2)}$ be two distinct thermal states of a given quantum system
corresponding to different temperatures $T_1$ and $T_2$ with $T_2>T_1$. We have
\begin{eqnarray}
-k_BT_1S(\rho^{(1)}||\rho_{*}) &=& -k_BT_1\ln{N}+\Omega_1-\Omega_{1}^{\prime},\label{eq:delta_1}\\
-k_BT_2S(\rho^{(2)}||\rho_{*}) &=& -k_BT_2\ln{N}+\Omega_2-\Omega_{2}^{\prime},\label{eq:delta_2}
\end{eqnarray}
where $\Omega_i=-k_BT_i\ln{Z_{i}}$ and $\Omega_{i}^{\prime}=-k_B(2T_i)\ln{Z_{i}^{\prime}}$ with $i=1,2$ are the thermodynamical potentials. By subtracting Eq.~(\ref{eq:delta_2}) from Eq.~(\ref{eq:delta_1}), we obtain
\begin{equation}\label{eq:geo_free1}
k_BT_2S(\rho^{(2)}||\rho_{*})-k_BT_1S(\rho^{(1)}||\rho_{*})
=\ln{\bigg[\frac{(\cos^2{d_B(\rho^{(2)},\rho_{*}}))^{k_BT_2}}{(\cos^2{d_B(\rho^{(1)},\rho_{*}}))^{k_BT_1}}\bigg]} 
=-\Delta \Omega+\Delta\Omega^{\prime}+k_B\Delta T\ln{N},
\end{equation}
where $\Delta x=x_2-x_1$ and we use Eq.~(\ref{eq:renyi_bures}). Thus, any transformation between $\rho^{(1)}$ and $\rho^{(2)}$ that changes the thermodynamical free energy can equivalently be understood as the change in the relative position with respect to the maximum entropy state in the state space of density operators. 

Our result, Eq.~(\ref{eq:geo_free1}), is a general one in the sense that we do not restrict ourselves with systems that are described by certain types of Hamiltonians. Secondly, all the other coordinates that can be included to the theory are encoded to the free energy and to the distance function through density operator formalism. One can, of course, apply the geometrical properties of the distance function to Eq.~(\ref{eq:geo_free1}) to obtain a pool of equalities and inequalities, though this is not a prime motivation of or an immediate issue for the current contribution.

However, one rogue element $Z^{\prime}=\sum_{n}e^{-\beta E_{n}/2}$ remains without a satisfactory physical interpretation. To make its contribution to the theory more explicit, we take the square of $Z^{\prime}$ to obtain
\begin{equation}\label{eq:z_pr}
Z^{\prime2}=Z(1+2\cos{d_W})=NZ\cos^2{d_B}.
\end{equation}
If we take the natural logarithm of both sides of Eq.~(\ref{eq:z_pr}), we obtain Eq.~(\ref{eq:fund1_1}).

To acquire a better understanding of $Z^{\prime}$ and the corresponding free energy $\Omega^{\prime}$, let us first recall the quantum relative entropy $D(\rho||\sigma):=\tr(\rho\ln\rho-\rho\ln\sigma)$ defined for all $\rho,\sigma$ acting on a Hilbert space of dimension $N$ becomes
\begin{equation}\label{eq:relative}
D(\rho_{th}||\rho_{*})=\ln N-\ln Z-\beta U,
\end{equation}
for an arbitrary thermal state $\rho_{th}=e^{-\beta\mathrm{H}}/Z$ with $U:=\tr(\rho_{th}\mathrm{H})$ being the internal energy. By obtaining the expression for $\ln Z$ from Eq.~(\ref{eq:relative}) and by using Eq.~(\ref{eq:fund1_1}), we find
\begin{equation}\label{eq:free_prime}
\Omega^{\prime}=\frac{1}{\beta}D(\rho_{th}||\rho_{*})+U+k_{B}TS(\rho_{th}||\rho_{*})-2k_{B}T\ln N.
\end{equation}
If we now consider $\rho^{(1)}$ and $\rho^{(2)}$ as the two distinct thermal states of a given quantum system
corresponding to different temperatures $T_1$ and $T_2$ with $T_2>T_1$ like before and calculate $\Delta\Omega^{\prime}$, Eq.~(\ref{eq:geo_free1}) gives
\begin{equation}\label{eq:trivial}
-\Delta\Omega+\Delta U = \frac{1}{\beta^{(1)}}D(\rho^{(1)}||\rho_{*})-\frac{1}{\beta^{(2)}}D(\rho^{(2)}||\rho_{*})+k_{B}\Delta T\ln N,
\end{equation}
which is trivial in the sense that we do not require the knowledge of Eq.~(\ref{eq:geo_free1}) to obtain it, though the inverse statement is not true. That is, by starting from Eq.~(\ref{eq:trivial}) one cannot obtain Eq.~(\ref{eq:geo_free1}) without the knowledge of the potential $\Omega^{\prime}$~(\ref{eq:free_prime}). A final straightforward calculation also yields $-\Delta\Omega+\Delta U=k_{B}T_2S(\rho^{(2)})-k_{B}T_1S(\rho^{(1)})$ with $S(\rho):=-\tr\rho\ln\rho$ being the von Neumann entropy as the equivalent form of Eq.~~(\ref{eq:trivial}), i.e., we recover the conventional thermodynamics. 

We complete this section by noting that
\begin{equation}\label{eq:occup_prime}
p_{i}=\frac{Ne^{-\beta E_{i}}}{Z^{\prime2}}e^{-S(\rho_{th}||\rho_{*})},
\end{equation}
for a given thermal state $\rho_{th}$. It follows that
\begin{equation}
\ln{p_{i}}=\ln{N}-\beta E_{i}-S(\rho_{th}||\rho_{*})-2\ln{Z^{\prime}},
\end{equation}
and thus,
\begin{equation}\label{eq:ent_ren_1}
TS_{th}-k_{B}TS(\rho_{th}||\rho_{*})=U+2k_{B}T\ln{Z^{\prime}}-k_{B}T\ln{N},
\end{equation}
where we write $S_{th}:=-k_{B}\sum_{i}p_{i}\ln{p_{i}}$ as the usual thermodynamical entropy. Now,
\begin{equation}
\ln{Z^{\prime}}=\ln{\sum_{i=1}^{N}e^{-\beta E_{i}/2}}=\ln{\sum_{i=1}^{N}p_{i}\bigg(\frac{1}{p_i}e^{-\beta E_{i}/2}\bigg)}\leq\sum_{i=1}^{N}p_{i}\ln{\bigg(\frac{1}{p_i}e^{-\beta E_{i}/2}\bigg)},
\end{equation}
by using Jensen's inequality. Therefore, we obtain
\begin{equation}\label{eq:ent_ren_2}
2k_{B}T\ln{Z^{\prime}}\leq 2TS_{th}-U.
\end{equation}
It is easy to see that the above inequality constitutes a first law for the auxillary potential $\Omega^{\prime}$ as it can be equivalently be written as $\Omega^{\prime}\geq U-2TS_{th}$. Finally, we combine Eqs.~(\ref{eq:ent_ren_1}) and~(\ref{eq:ent_ren_2}) to give
\begin{equation}\label{eq:ent_bound}
S_{th}\geq k_{B}\big(\ln{N}-S(\rho_{th}||\rho_{*})\big).
\end{equation}
The presented geometric bound to the equilibrium entropy $S_{th}$ can more easly be obtained via an application of Jensen's inequality to our fundamental equation~(\ref{eq:fund2}) at the expense of the knowledge gained from Eq.~(\ref{eq:occup_prime}) to Eq.~(\ref{eq:ent_ren_2}). 
\subsection{Work and distance}
Before we proceed, let us define $S_{R}^{(i)}=S(\rho^{(i)}||\rho_{*})$ and $\Delta_{T}x:=k_{B}T_{2}x^{(2)}-k_{B}T_{1}x^{(1)}$ for the sake of clearity. It follows from Eq.~(\ref{eq:geo_free1}) that
\begin{equation}
-\Delta\Omega=\Delta_{T}S_{R}-k_{B}\Delta T\ln{N}-\Delta\Omega^{\prime}\geq W
\end{equation}
where $W\geq0$ is the positive work done by the sytem during the transformation. Rearranging the terms such that $W+\Delta\Omega^{\prime}\leq\Delta_{T}S_{R}-k_{B}\Delta T\ln{N}$ and due to the facts that $\Delta_{T}S_{R}-k_{B}\Delta T\ln{N}\leq0$ and $W\geq0$, we obtain
\begin{equation}\label{eq:work1}
W\geq\Delta_{T}S_{R}-k_{B}\Delta T\ln{N},
\end{equation}
or equivalently,
\begin{eqnarray}
\label{eq:carnot_work_distance1}
\frac{\eta_{c}}{1-\eta_{c}}&\geq&\frac{\Delta_{T}S_{R}-W}{k_{B}T_1\ln{N}},\\
\label{eq:carnot_work_distance2}
\eta_{c}&\geq&\frac{\Delta_{T}S_{R}-W}{k_{B}T_2\ln{N}},
\end{eqnarray}
where $\eta_{c}:=1-(T_{1}/T_{2})$ is the Carnot efficiency. 

A more general relation between the moments of work and distance can be obtained as follows. We racall that
\begin{equation}
S(\rho_{th}||\rho_{*})=\ln{N}+\ln{Z}-2\ln{Z^{\prime}}=\ln{N}-\ln{Z}+\ln{\bigg(\frac{Z^2}{Z^{\prime2}}\bigg)}.
\end{equation}
By expanding the last term $\ln{(Z^2/Z^{\prime2})}$ up to its first order, we write
\begin{eqnarray}
\nonumber k_{B}TS(\rho_{th}||\rho_{*})&\sim& k_{B}T\ln{N}+\Omega+\frac{k_{B}T}{N}e^{-\beta\Omega}e^{S(\rho_{th}||\rho_{*})}-k_{B}T,\\
&\sim&k_{B}T\ln{N}+\Omega+\frac{k_{B}T}{N}e^{S(\rho_{th}||\rho_{*})}(1-\beta\Omega)-k_{B}T,
\end{eqnarray}
where $T$ is the equilibrium temperature of the system. By rearranging the terms and solving for the potential $\Omega$, we obtain
\begin{equation}\label{eq:first_order}
\Omega^{(1)}\sim k_{B}T\frac{S(\rho_{th}||\rho_{*})-\frac{1}{N}e^{S(\rho_{th}||\rho_{*})}-\ln{N}+1}{1-\frac{1}{N}e^{S(\rho_{th}||\rho_{*})}},
\end{equation}
as the first order approximation to the free energy. Similarly, the contribution of the second order term from the expansion of $\ln{(Z^2/Z^{\prime2})}$ leads
\begin{equation}\label{eq:second_order}
\Omega^{(2)}\sim k_{B}T\frac{S(\rho_{th}||\rho_{*})-\frac{2}{N}e^{S(\rho_{th}||\rho_{*})}+\frac{1}{2N^2}e^{2S(\rho_{th}||\rho_{*})}-\ln{N}+\frac{3}{2}}{1-\frac{2}{N}e^{S(\rho_{th}||\rho_{*})}+\frac{1}{N^2}e^{2S(\rho_{th}||\rho_{*})}}.
\end{equation}
It is easy to recognize the pattern
\begin{equation}
1-\frac{1}{N}e^x+\frac{1}{N^2}e^{2x}-...\sim\bigg(1+g\frac{e^x}{N}\bigg)^{-1/2},
\end{equation}
with $g$ being a constant. Thus, we obtain
\begin{equation}\label{eq:free_energy_approx}
\Omega\sim k_{B}T\bigg[\sqrt{1+g\frac{e^{S(\rho_{th}||\rho_{*})}}{N}}\bigg(S(\rho_{th}||\rho_{*})-\ln{N}+\frac{1}{2}\bigg)+1\bigg],
\end{equation}
as the approximate geometric desciption of the thermodynamical free energy.

As before, let $\rho^{(1)}$ and $\rho^{(2)}$ be two distinct thermal states of a given quantum system
corresponding to different temperatures $T_1$ and $T_2$ with $T_2>T_1$. Let us define 
\begin{equation} h^{(i)}:=\sqrt{1+g\frac{e^{S_{R}^{(i)}}}{N}}\bigg(S_{R}^{(i)}-\ln{N}+\frac{1}{2}\bigg),
\end{equation}
where $i=1,2$. Thus, for any thermodynamical transformation $\rho^{(1)}\rightarrow\rho^{(2)}$, the change in the free energy reads,
\begin{equation}\label{eq:geo_free_chng}
-\Delta\Omega\sim-\Delta_{T}h-k_{B}\Delta T.
\end{equation}
More, in terms of the Carnot efficiency, we obtain
\begin{equation}\label{eq:geo_free_chng_crnt}
\eta_{c}\sim\frac{1}{k_{B}T_2}(\Delta\Omega-\Delta_{T}h).
\end{equation}
For our final touch, we use the Jarzynski equality~\cite{jarzynski1997nonequilibrium} to write,
\begin{equation}\label{eq:work_distance}
\frac{\eta_c}{1-\eta_c}\sim-\frac{1}{k_{B}T_1}\Delta_{T}h-\ln{\bigg\langle\exp{\frac{-W}{k_{B}T_1}}\bigg\rangle},
\end{equation}
where $\langle ,\rangle$ denotes the average over an ensemble of measurements performed on the work $W$. We read the Eq.~(\ref{eq:work_distance}) as: during a thermodynamical transformation between two quantum thermal states, maximum work can be extracted with maximum efficiency by optimizing the difference of the relative distances of each state with respect to the maximum entropy state while keeping all the other parameters (forces) constant. 

Eq.~(\ref{eq:work_distance}) can particularly be useful for the investigations of systems that undergo quantum phase transitions (QPTs), as the Bures distance and the moments of work considered to be witnesses of QPTs, separately~\cite{zanardi2007bures,zanardi2007information,you2007fidelity,francica2016driven}. This connection requires further calculations that are out of the context and can be a topic of another contribution. 
\begin{figure}[!t]
	\centering
	\subfigure[]{
		\label{fig:fig1a}
		\includegraphics[width=6cm]{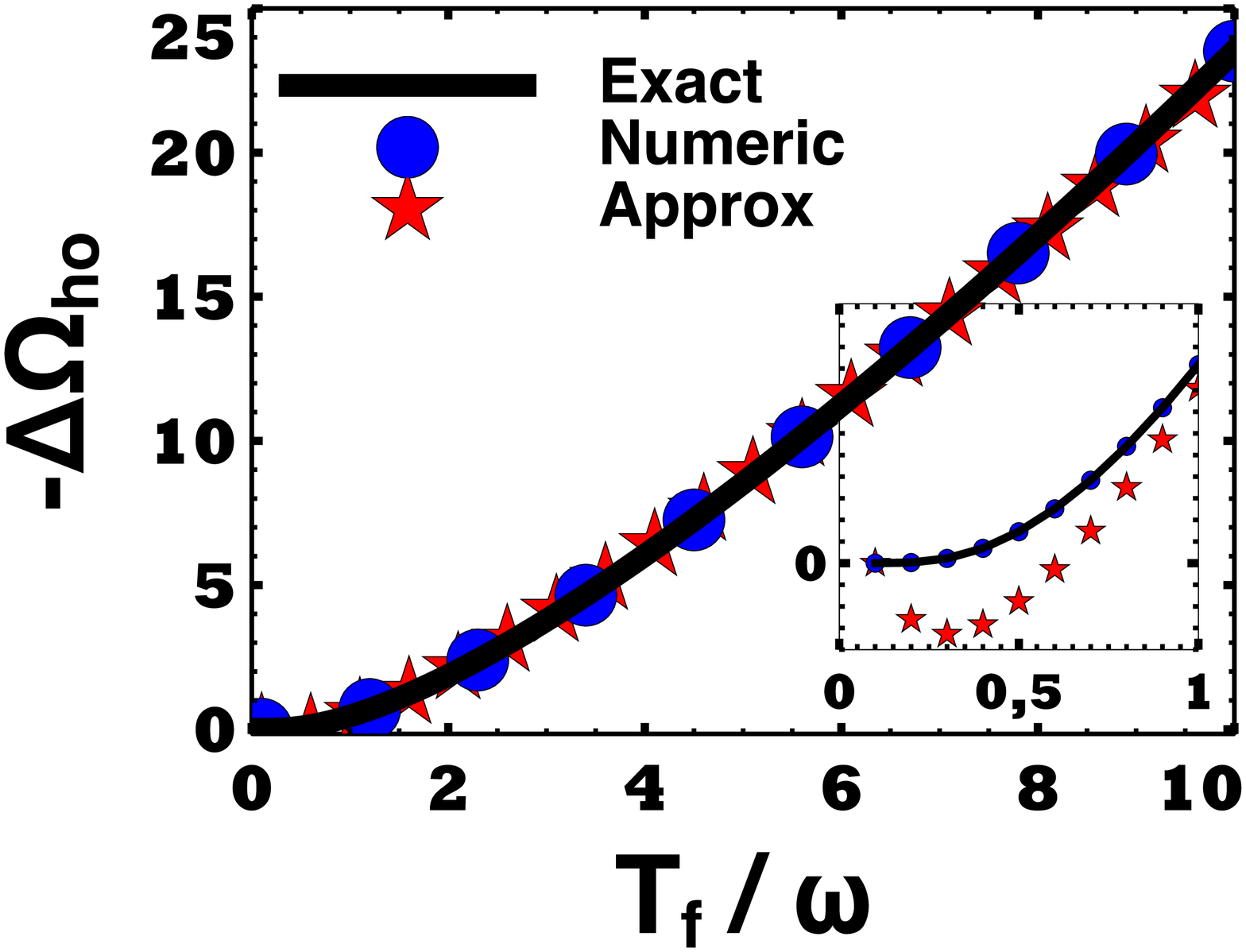}
	}
	\subfigure[]{
		\label{fig:fig1b}
		\includegraphics[width=6cm]{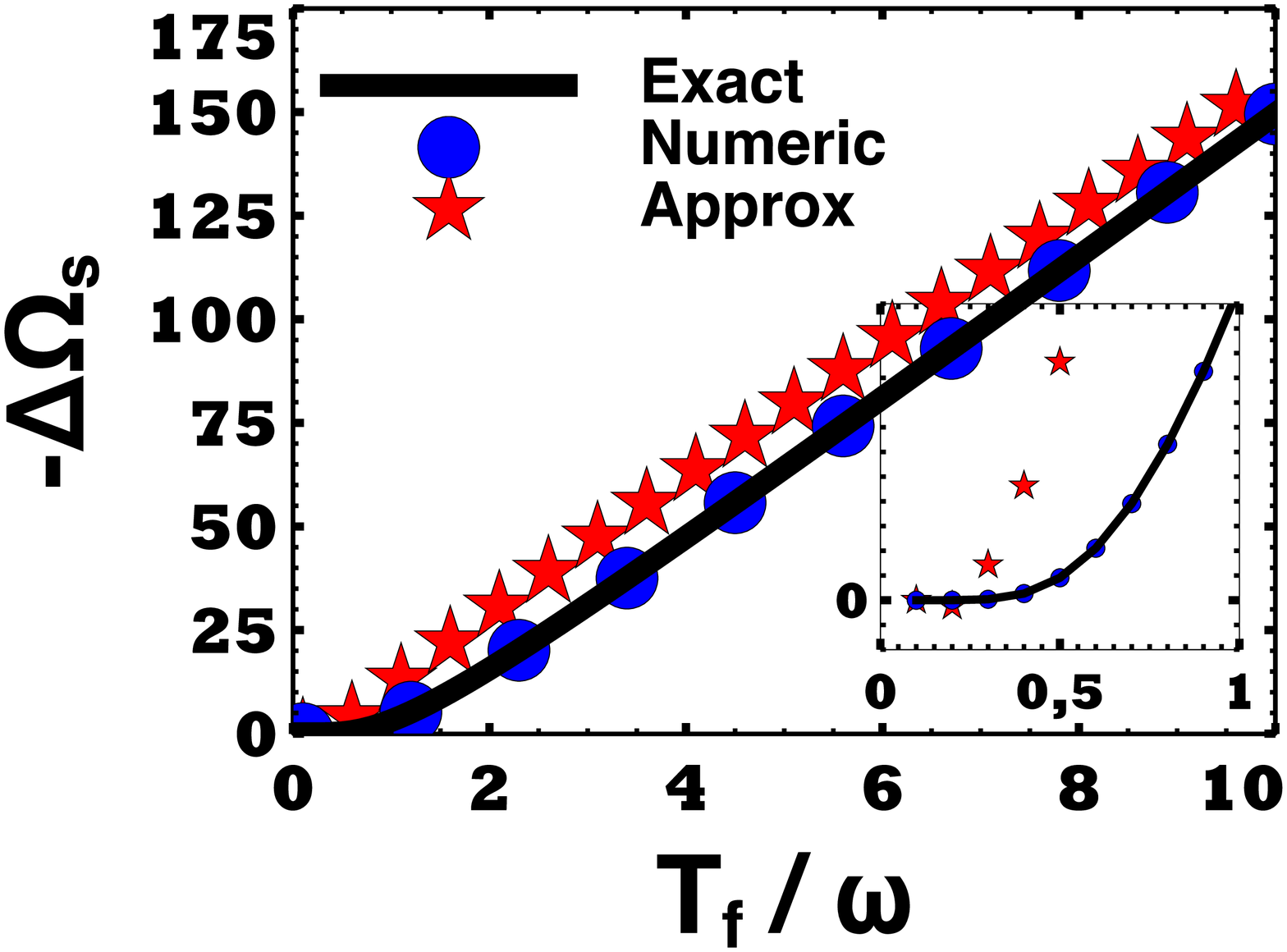}
	}
	\caption{
		Free energy changes for (a) quantum harmonic oscillator with $N=100$ and (b) an ensemble of $n_s=25$ ($N=2^{n_s}$) spin-$1/2$ particles with respect to the scaled temperature $T/\omega$. Black-solid line represents the exact results obtained via claculating $-\Delta\Omega=T_f\ln{Z_f}-T_i\ln{Z_i}$, while blue-circles and red-stars represents the data obtained from~Eq. (\ref{eq:geo_free1}) and Eq.~(\ref{eq:geo_free_chng}) with $g=\pi$, respectively. All the other parameters are as explained in the text.  
	}
	\label{fig:fig1}
\end{figure}
\subsection{Examples}
We are now ready to complete our analysis with some numerical examples. To that end, we consider two fundamental quantum systems, quantum harmonic oscillator and an ensemble of spin $1/2$ particles, described by the Hamiltonians ($\hbar=1$) $\mathrm{H}_{ho}=\omega_f\hat{a}^{\dagger}\hat{a}$ and $\mathrm{H}_{s}=\omega_a\hat{S}_z$, respectively. Here, $\hat{a}^{\dagger}$ $(\hat{a})$ and $\hat{S}_z=(1/2)\sum_{k}\hat{\sigma}_z^{k}$ are the bosonic creation (annihilation) and collective spin operators and we denote resonance frequencies with $\omega_f$ and $\omega_a$, respectively.

We assume that the both systems are in thermal equilibrium at temperature $T_{i}/\omega=0.1$ ($k_B=\hbar=1$, $\omega_f=\omega_a=\omega$). We calculate the change in thermodynamical free energy, $-\Delta\Omega=T_f\ln{Z_f}-T_i\ln{Z_i}$, with respect to final temperature $T_f$ and compare it with the numerical data obtained from Eq.~(\ref{eq:geo_free1}) and Eq.~(\ref{eq:geo_free_chng}) where we set $g=\pi$ for both systems. 

In Fig.~\ref{fig:fig1a}, we show the free energy change in quantum harmonic oscillator for $N=100$. Eq.~(\ref{eq:geo_free1}) is in perfect agreement with the exact results. Our approximated result, Eq.~(\ref{eq:geo_free_chng}), agrees well within the temperature regime except for a small anomaly occuring in the low temperatures as shown in the inset. We numerically verified that for even higher dimensions (i.e., $N>100$), the offset minimalizes. 

Fig.~\ref{fig:fig1b} depicts the change in free energy for an ensemble of $n_s=25$ ($N=2^{n_s}$) spin-$1/2$ atoms. As in the bosonic case Eq.~(\ref{eq:geo_free1}) is in perfect agreement with the analytical results. At this point we also note that Eq.~(\ref{eq:geo_free1}) gives precise results for a single two-level system, as well. The approximation diverges greater than that of we calculate for the bosonic case in the temperature regime under consideration.

To test the exact bounds on positive work done by a system, i.e., Eq.~(\ref{eq:carnot_work_distance1}), we consider generalized quantum Rabi model ($\hbar=1$)
\begin{equation}\label{eq:rabi}
\hat{H}=\frac{\omega}{2}\hat{\sigma}_z+\epsilon\hat{\sigma}_x+\omega \hat{a}^{\dagger}\hat{a}+g\hat{\sigma}_x(\hat{a}+\hat{a}^{\dagger})
\end{equation}
as a hybrid spin-boson quantum Otto engine. The model has already been considered in the literature and we refer to Ref.~\cite{altintas2015rabi} for details. Here, $\omega$ is the resonance frequency of the system, $g$ is the strength of atom-field coupling and $\epsilon$ is a small coefficient which breaks the $Z_2$ symmetry of the model. We assume that the engine operates between the temperatures $T_1 = 0.05$ and $T_2\in\{0.2, 0.25\}$ (scaled with $\hbar\omega/\kappa_B$) with the corresponding frequencies of $\omega_1 = \omega$ and $\omega_2 = 2\omega$ and we set $\epsilon=0.005\omega$.

\begin{figure}[!t]
	\centering
	\subfigure[]{
		\label{fig:fig2a}
		\includegraphics[width=6cm]{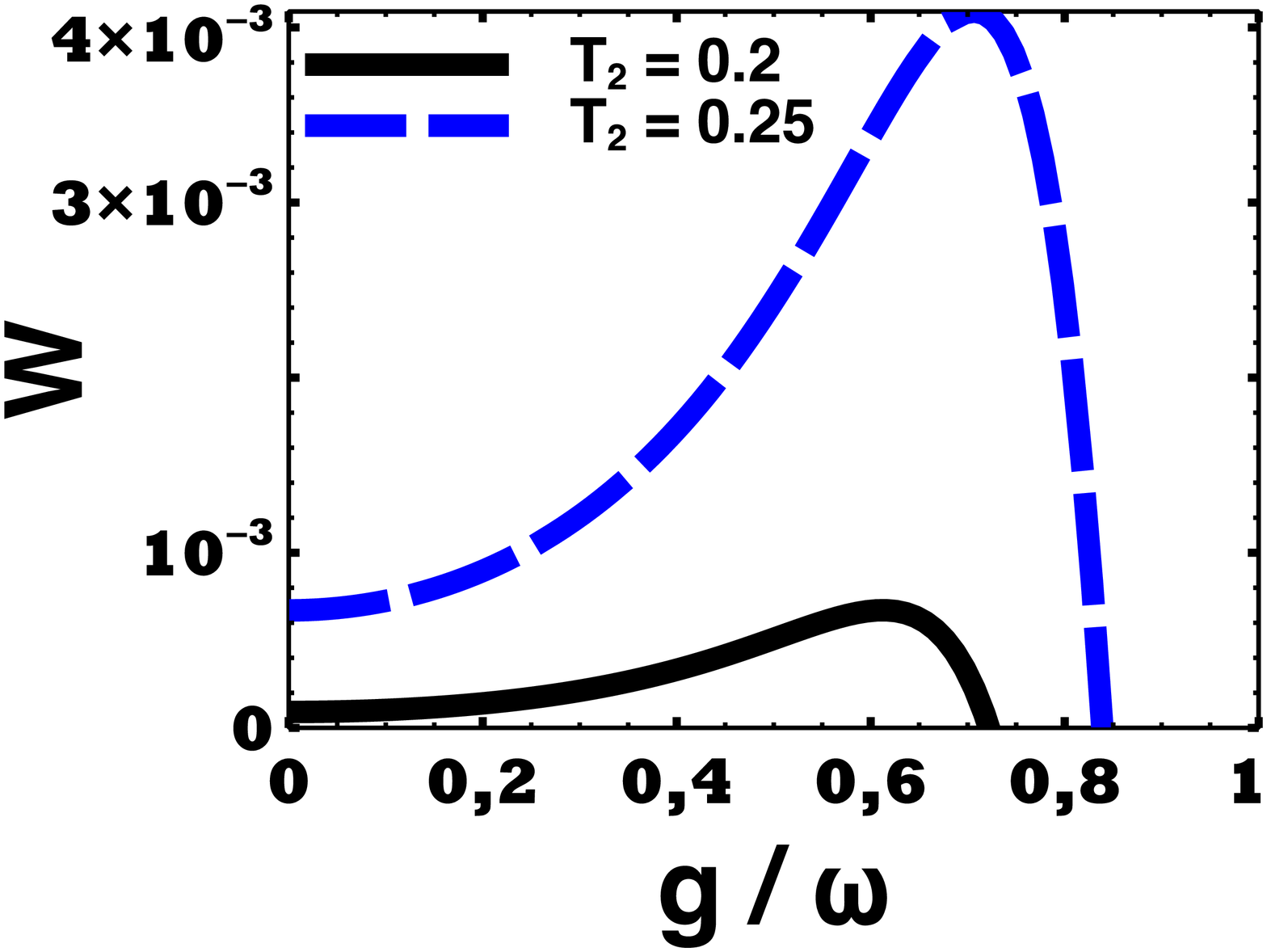}
	}
	\subfigure[]{
		\label{fig:fig2b}
		\includegraphics[width=6cm]{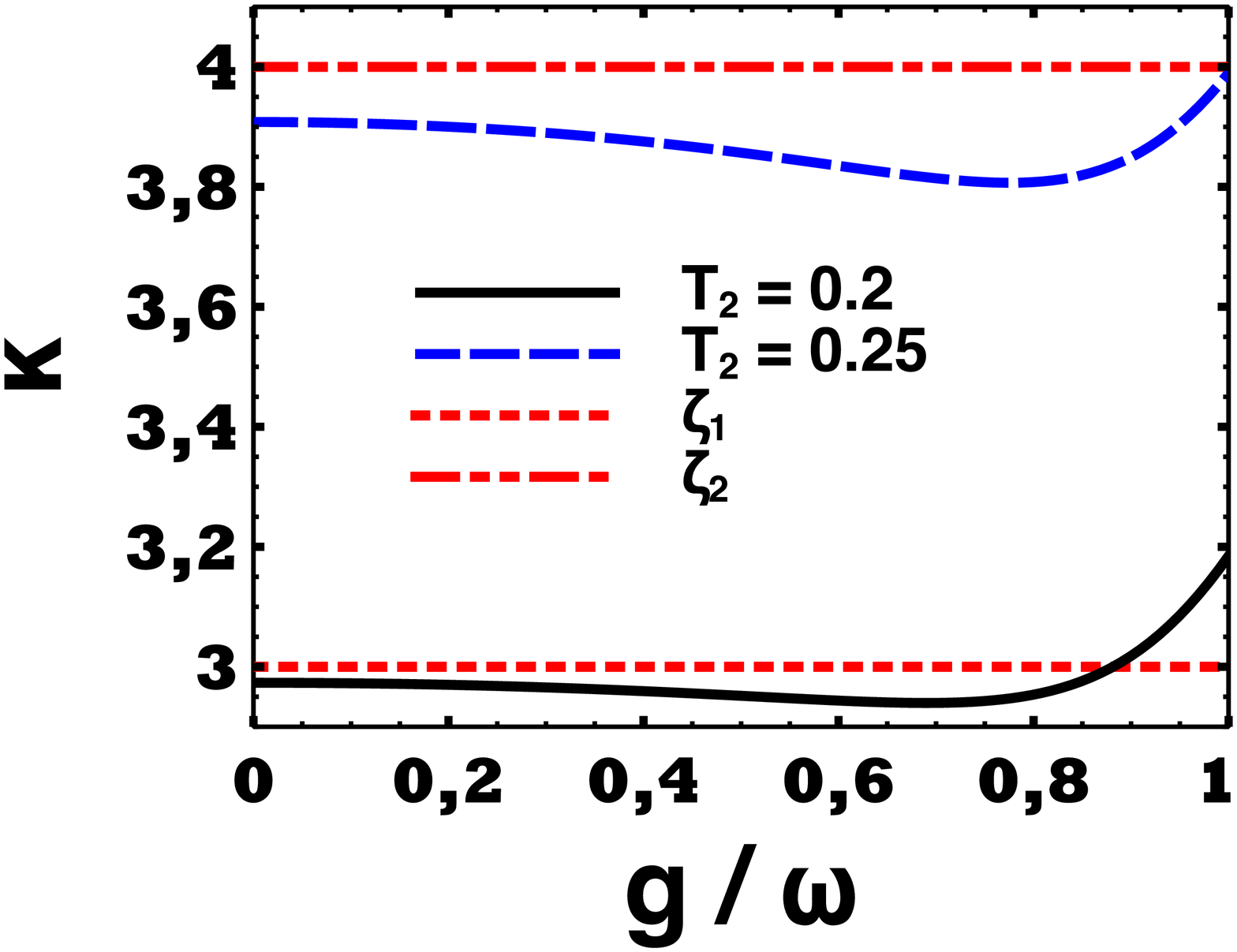}
	}
	\caption{
		(a) Extracted work from Rabi system as a quantum Otto engine for $T_2=0.2$ (black-solid) and $T_2=0.25$ (blue-dashed) (b) The change in the parameter $\kappa=(\Delta_{T}S_{R}-W)/(T_1\ln{N})$ for $T_2=0.2$ (black-solid) and $T_2=0.25$ (blue-dashed). Corresponding upper bounds $\zeta_1=3$ and $\zeta_2=4$ are flagged with red-dotted and red dot-dot-dashed lines. x-axis is the scaled interaction strength $g/\omega$ in both figures. All the other parameters are as explained in the text.  
	}
	\label{fig:fig2}
\end{figure}

Let us define $\zeta:=\eta_c/(1-\eta_c)$ and $\kappa:=(\Delta_{T}S_{R}-W)/(T_1\ln{N})$ with $N = n_{boson}n_{spin}=30\times2=60$. It follows that we have $\zeta_1=3$ and $\zeta_2=4$ for $T_2=0.2$ and $T_2=0.25$, respectively. Fig.~\ref{fig:fig2} shows our typical results for $\kappa$ with respect to $g/\omega$ with the verification of analytical results.
\section{Concluding remarks}\label{sec:conc}
All of our results concerning geometry and quantum statistics follow from distance between two quantum equilibrium states. They are not explicit in the conventional theory and requires relative measurements with respect to the maximally mixed state to surface out.

The use of density operator formalism in the construction of geometry and statistical thermodynamics led to a general, system and process independent theory. Furthermore, all of the relations rise as the functions of the occupation probabilities instead of the thermal entropy, $S_{th}$, itself. The latter, as a fundamental requirement for statistical explorations of pysical systems within the quantum mechanical framework, bring consistency to the theory.

Finally, the leading results concerning the ties among efficiency, distance, work and its moments, as being established in equilibrium, present new angles to the emerging field of thermal quantum machines. They simply suggest distance optimization procedures to have robust work harvesting with maximum possible efficiency. Another particular implication is that our results combine, and verify seperate discussions on the detection of QPTs via either Bures geometry~\cite{zanardi2007bures,zanardi2007information,you2007fidelity} or the moments of work~\cite{francica2016driven}. Indeed, as their difference is bounded by or equalised to a universal constant, then, if one of them can detect a physical phenomena, the detection with the same event by other is inevitable.

\vspace{6pt} 

\acknowledgments{We thank  M. \"{O}zkan and J. Vaccaro for illuminating discussions. A. \"{U}. C. H. acknowledges the COST Action MP1209.  A. \"{U}. C. H. and \"{O}. E. M. acknowledge the support from Koc University and Lockheed Martin Corporation Research Agreement.}

\authorcontributions{A. \"{U}. C. H. conceived the idea and derived the technical results. A. \"{U}. C. H. and \"{O}. E. M. developed the theory and wrote the manuscript together.}

\conflictofinterests{The authors declare no conflict of interest.}
\bibliographystyle{mdpi}
\bibliography{geo_thermo}
\end{document}